Version 15 July 1994

Angular Dependent Soft X-ray Absorption Spectroscopy of $V_2O_5$ and $V_6O_{13}$

E. Goering,[a] O. Müller,[a] M. Klemm,[a] J. P. Urbach,[a] H. Petersen,[b] C. Jung,[b] M. L. denBoer,[c] and S. Horn[a]

[a]Institut für Physik, University of Augsburg, 86159 Augsburg, Germany

[b]BESSY, Lentzealle 100, 14195 Berlin, Germany

[c]Physics Dept., Hunter College CUNY, New York NY 10021, USA

We have measured the V $2p_{3/2}$ and O $1s$ x-ray absorption spectra of single crystal $V_2O_5$ and $V_6O_{13}$ and compared to linear combinations of atomic orbitals (LCAO) density of states (DOS) calculations. The spectra change dramatically with incident angle. The use of polarized light and a single crystal limits the number of transitions possible, revealing spectral features that cannot be resolved on polycrystals (angle-integrated). The measured O $1s$ and V $2p_{3/2}$ spectra agree with the projected unoccupied O $2p$ and V $3d$ DOS respectively, indicating that atomic and solid state effects, including V-O hybridization, must be included to adequately describe the spectra.

Corresponding author address:    Eberhard Goering

Memminger Str. 6

86159 Augsburg

Fax   +49 821 5977 411

email   ebs@physik.uni-augsburg.de

Near-edge x-ray absorption spectra (NEXAFS) of condensed matter systems, particularly narrow-band transition metal (TM) based systems such as high temperature superconductors, vanadium oxides and oxide magnets, are widely studied. This activity has been spurred by theoretical and experimental progress. However, the very important *d* states in TM-based systems remain difficult; intraatomic correlations and solid state effects give rise to complicated spectra. We show here linearly polarized light on single crystals of anisotropic materials mitigates these difficulties and provides a powerful probe of electronic state properties.

We have studied the V 2*p* and O 1*s* spectra of $V_2O_5$ and $V_6O_{13}$. The former is a layered compound with orthorhombic Pmmn (a = 11.51 Å, b = 3.563 Å, c = 4.369 Å) symmetry. The vanadium atom is surrounded by six oxygen atoms that form a distorted octahedron. One V - O distance is 1.585 Å while the distance to O in adjacent layers is 2.785 Å. The layered structure means that $V_2O_5$ should show particularly sharp anisotropic absorption spectra. $V_6O_{13}$ is monoclinic C2/m (a = 11.92 Å, b = 3.68 Å, c = 10.14 Å, β = 100.87°).

Single crystals of $V_2O_5$ were prepared by heating a $V_2O_5$ powder in air above the melting temperature and then slowly cooling through the liquid-solid transition. $V_6O_{13}$ was prepared by chemical transport using $TeCl_4$. Spectra were measured at BESSY I in Berlin using the new HE-PGM3 monochromator which features resolution up to 10,000. Samples were cleaved in the *ab* plane in ultra-high vacuum, in which they were found to be very stable. Spectra were taken by measuring the total electron yield of the sample and were normalized to the incident flux and to the combined increase in absorption at the V 2*p* and O 1*s* edges.

Fig. 1 shows the V $2p_{3/2}$ and O 1*s* x-ray absorption spectra of $V_2O_5$ at various angles Φ between the polarization vector *E* of the incident x-rays and the *c* axis of the $V_2O_5$ crystal. Dramatic changes are apparent as a function of this angle, while rotating *E* within the *ab* plane, i.e. perpendicular to the *c* axis, causes no changes (not shown). A polycrystalline sample, being essentially an average over all orientations, would produce far less detailed information.

The angular dependence of the spectra allows qualitative determination of the orientation of the orbitals to which spectral features correspond. For a quantitative description, a model for the unoccupied states is required. The observed strong angular dependence and the narrowness of the bands near $\epsilon_F$ suggest a linear combination of atomic orbitals (LCAO) description of the hybridized O 2*p* and V 3*d* states.[1] Surprisingly, final state effects appear minimal, even for the V $2p_{3/2}$ edge.[1] Our LCAO calculations used the CRYSTAL code [2] which incorporates *ab initio* Hartree-Fock equations for the periodic $V_2O_5$ lattice. The calculated lowest energy unoccupied states are formed primarily by antibonding hybridized V 3*d* and O 2*p* orbitals, with some contributions from higher-lying orbitals such as V 4*sp*, O 3*sp* and O 4*sp*. The structure of the V 2*p* (O 1*s*) edge therefore reflects largely the V 3*d* (O 2*p*) contribution to these antibonding states.

In Fig. 2 the results of our LCAO calculation, broadened to account for lifetime effects and experimental resolution, are compared to the measured spectra. The O1*s* spectrum at $\Phi = 0°$ (Fig. 2b), where dipole transitions to O *p* orbitals located in the *ab* plane are allowed, is in good agreement with the calculated projected unoccupied density of states based on O $p_x$ and $p_y$ orbitals, (*x*, *y*, and *z* are parallel to the *a*, *b*, and *c* crystal axes, respectively) considering the neglect of matrix element and final state effects. Both curves show two prominent peaks about 2.5 eV apart, to which the $2p_x$ and $2p_y$ orbitals contribute equally in agreement with the experimental observation that the spectrum does not change when ***E*** is rotated within the ab plane. Further support is provided by the $\Phi$ dependence. When ***E*** is oriented almost perpendicular to the ab plane ($\Phi = 85°$, grazing incidence, Fig. 2b), so that dipole transitions are only allowed into empty states of oxygen $2p_z$ symmetry, the measured spectrum resembles the projected $p_z$ DOS. At this angle both measurement and calculation have greater intensity at higher energy while the low energy portion has decreased substantially.

We now consider the V 2$p$ spectrum, where intraatomic Coulomb interactions may play a more important role. The energy separation of the V $2p_{1/2}$ and $2p_{3/2}$ edges prevent overlap [1] and it suffices to consider the $2p_{3/2}$ edge. Fig. 2a shows the projected V 3$d$ unoccupied DOS corresponding to orbitals in the ab plane, to which orbitals of $d_{xy}$, $d_{x2-y2}$, $d_{xz}$, and $d_{yz}$ character contribute, and the out-of-plane component derived from orbitals of $d_{3z2-r2}$, $d_{xz}$, and $d_{yz}$ character, broadened as for oxygen. The in-plane and out-of-plane components differ most below 518 eV, where the in-plane component displays a prominent peak (largely $x^2$-$y^2$ and $xy$ in origin) absent in the out-of-plane component, which has more intensity near 520 eV. These orientational differences are similar to those in the measured absorption spectra (Fig. 2a): the two low energy peaks seen for ***E*** parallel to the *ab* plane decrease when ***E*** is turned out-of-plane and a peak develops at slightly higher energy.

In $V_6O_{13}$ our measurements indicate that the $\Phi$ dependence of the V $2p_{3/2}$ spectra (not shown) is not as dramatic as for $V_2O_5$, although the general trend is similar. On the other hand, the O 1$s$ spectra are a function of both $\Phi$ and $\theta$, as illustrated in Fig. 3. The $\theta$ dependence implies a stronger anisotropy within the *ab* plane than for $V_2O_5$.

In summary, we have shown that, on low symmetry crystals, angle resolved absorption measurements can provide detailed information due to the selectivity afforded by the use of linearly polarized light on single crystal samples. This suggests that x-ray absorption spectroscopy may be capable of verifying in great detail theoretical descriptions of the electronic structure of transition metals oxides, at least when highly directional bonds are present. Quantitative comparison with theory appears possible.

We appreciate funding from the German BMFT, Förderkennzeichen 05 5WAABB 0.

**Figure captions**

Fig. 1  Dependence of the x-ray absorption spectrum of $V_2O_5$ on the angle $\Phi$ (see text) in the vicinity of the V $2p_{3/2}$ edge and the O $1s$ edge.

Fig. 2  (a) The O $1s$ edge measured at $\Phi = 0°$ compared to the broadened calculated projected DOS originating from in-plane components of O $2p$ orbitals, and the O $1s$ edge measured at $\Phi = 80°$ compared to the DOS originating from out-of-plane components of O $2p$ orbitals.

(b) The V $2p_{3/2}$ edge measured at $\Phi = 0°$ compared to the broadened calculated projected DOS originating from in-plane components of V $3d$ orbitals, and the V $2p_{3/2}$ edge measured at $\Phi = 80°$ compared to the DOS originating from out-of-plane components of V $3d$ orbitals.

Fig. 3  Dependence of the O $1s$ spectrum of $V_6O_{13}$ on $\Phi$ at $\theta = 0°$ and $\theta = 90°$.

Fig.1

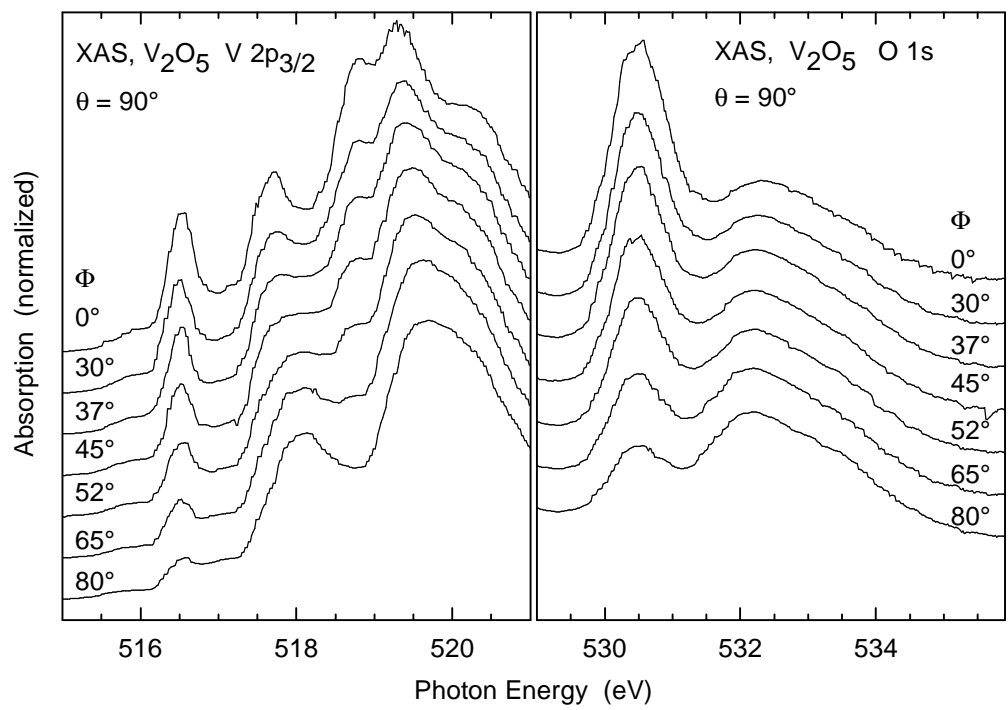

Fig.2

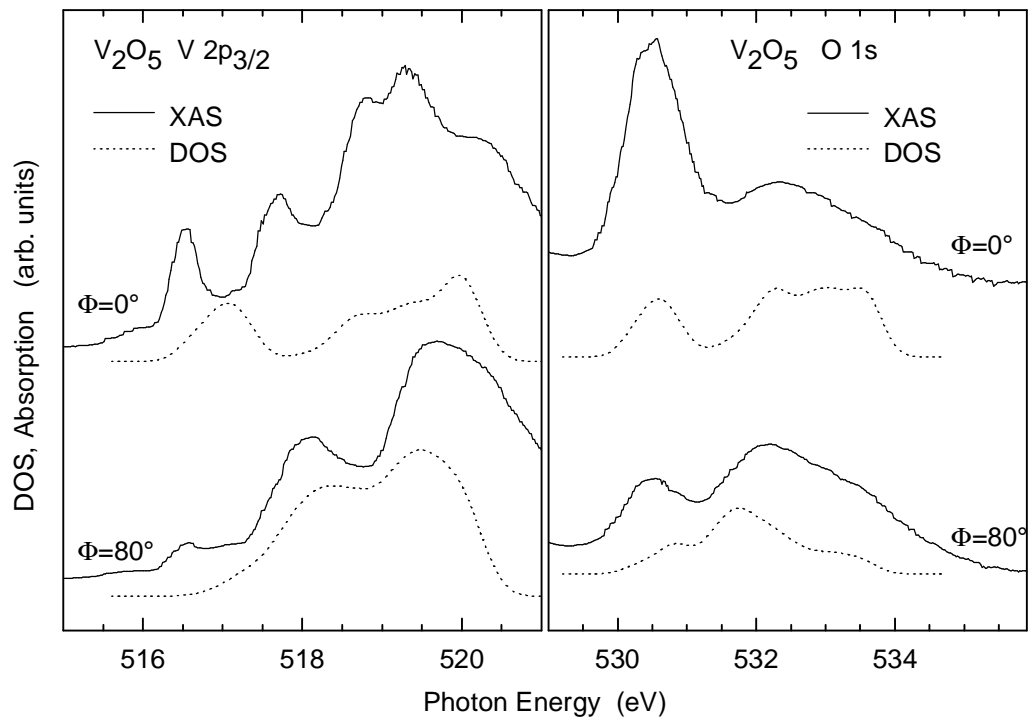

Fig.3

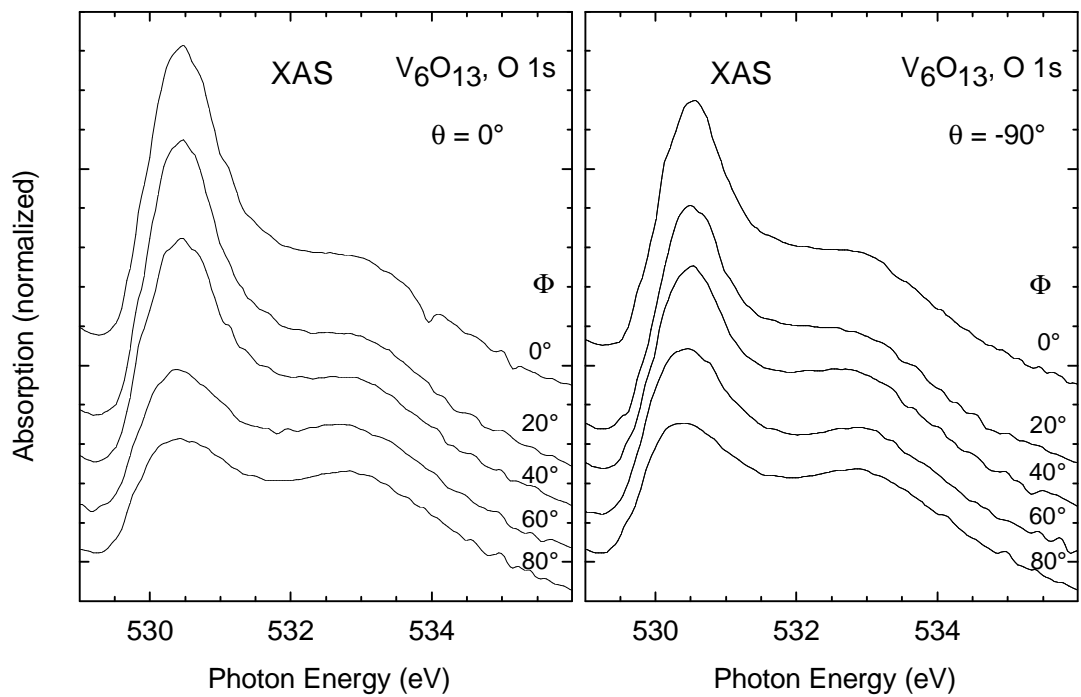